\documentclass[twocolumn,floatfix,aps,showkeys,showpacs]{revtex4}

\usepackage{amsmath}
\usepackage{amssymb}
\usepackage{graphicx}
\usepackage{enumerate}

\begin{document}

\title{Bias in generation of random graphs}

\author{Hendrike Klein-Hennig}
\author{Alexander K. Hartmann}
\email{a.hartmann@uni-oldenburg.de}
\affiliation{Institute of Physics, University of Oldenburg, 26111 Oldenburg, 
Germany}

\date{\today}

\begin{abstract}
We study the statistical properties of the 
generation of random graphs according the configuration
model, where one assigns randomly degrees to nodes. This model
is often used, e.g., for
 the scale-free degree distribution $\sim d^\gamma$.
For the efficient variant, where non-feasible edges are rejected
and the construction of a graph continues,
there exists a bias, which we calculate explicitly for
a small sample ensemble. We find that this 
bias does not disappear with growing system
size. This becomes also visible, e.g., for scale-free
graphs when  measuring quantities like
the graph diameter.  Hence, the
efficient generation of general scale-free graphs with a very broad
distribution ($\gamma <2$) remains an open problem. 
\end{abstract}


\maketitle

�\section{Introduction} 

Networks have become a very valuable tool when analyzing
complex systems like social communities, protein interactions,
the Internet or the spread of diseases 
\cite{watts1999b,albert2002review,dorogovtsev2003,newman2003review,%
boccaletti2006,newman2006b}.
There are two basic approaches to analyze the
creation, structure,  and behavior of networks: One is to look at 
specific real-world networks and analyze as many parameters as
possible, comparing them to other specific real-world networks. The
second approach is to generalize form the given data
and to find  network ensembles which describe one or several  real-world
networks as close as possible.
These ensembles are intended to be generated within computer simulations
\cite{practical_guide2009} and analyzed  using statistical methods. Thus, 
one has to find methods for generating model networks exhibiting the
desired features. These samples should have good statistical
properties, which means that each realization of the graph should be
created with a desired probability, often this 
the uniform ensemble. A generation method which fulfills
this is called \emph{unbiased}.

Well known ensembles are \emph{small-world} networks \cite{watts1998,watts1999a}
and scale-free networks, the latter exhibiting a power-law
distribution with density
\begin{equation}
P(d)\sim d^{-\gamma} \quad (d>0)
\label{eq:scale-free}
\end{equation}
for the degrees $d$, i.e., the number
of neighbors of a node. 
Such a behavior for the degree distribution is often observed for real-world
systems \cite{redner1998,broder2000,barabasi2000,jeong2000,%
liljeros2001,newman2006}. 
A very efficient method to generate random graphs
is \emph{preferential attachment}
\cite{deSollaPrice1976,barabasi1999,dorogovtsev2000,krapivsky2001}.
For some networks, like citation networks, this is a very
suitable model. Nevertheless,
this does not allow to generate graphs with a very broad degree
distribution $\gamma <2$. Also,� 
the preferential attachment process does create certain
correlations, in particular the obtained graphs are always connected.
Hence, for general models which do not make any assumptions
beyond the degree distributions, other methods have to be used.

A more general approach is to first draw a degree sequence from
the desired degree distribution and in a second step to assign
the edges randomly such that all simple 
graphs (without multiple connections or self-loops)
which are feasible for this degree sequence are equiprobable
\cite{molloy1995}. 
Note that this generates \emph{labeled} graphs, i.e., each node
is distinguishable from the other nodes. This means, e.g., the 
(single) graph
with degrees $d_1=2$, $d_2=1$, $d_3=1$ is different from the
graph exhibiting $d_1=1$, $d_2=2$, $d_3=1$.
A method which is frequently used for graphs with
predefined degree sequences is the \emph{configuration model}. Bender and
Canfield \cite{Bender1978} and  B\'ollobas \cite{bollobas1980}
introduced the mathematical background in 1978--1980. 
A very efficient algorithm was described 
by Newman et al. \cite{newman2001} in 2001. For
each vertex with a given degree, \emph{stubs}
are created, which are
the points where edges are emerging from the vertex. In a second
step random pairs of stubs are connected until there are no stubs
left. In order to be able to connect all stubs the total number of
stubs must be even. For randomly drawn sets of stubs this can be
archived by disregarding sets with an odd number of stubs and
generating a new set. In the configuration model, however, it is
possible that two stubs of the same vertex are connected creating a
self loop, or two different vertices are connected with multiple
edges. When statistically analyzing networks usually one
considers simple graphs. Therefore self-loops and multiple edges have to be
avoided when generating these graphs in a computer.
 There are several possibilities how to deal with this problem
described in \cite{Britton2006}.

 One method (``\emph{refusal}''),  
is to disregard all non-simple graphs and redo
the algorithm until a simple graph is created. In other words,
as soon as a self-loop
or multiple edge is created all connections made so far have to be
disregarded and the generation process is restarted. This procedure
will generate all graphs with a given set of degrees with equal
probability \cite{Britton2006}. The disadvantage of this procedure is
that many attempts may be needed to create a simple graph with this
method,. This becomes quite annoying 
in particular for scale free graphs with a broad degree
distribution ($\gamma <2$) where for a large number of nodes
it becomes impossible to generate a single graph instance.
For example, if one accepts for constructing a graph 
up to one CPU hour,  for $\gamma=1$ only graphs with about 
$n=30$ nodes are feasible  while for $\gamma=2$ one can go up to $n=300$.

In practice, in many publications a different approach (``\emph{repetition}'')
is used:
 Much fewer graphs are thrown away if, when encountering the
generation of a forbidden edge, the connections made so far are kept
and only the last connection is disregarded and a new pair of stubs
is randomly drawn as explicitly mentioned 
by Milo et al.\ \cite{Milo2002} or later on by 
Catanzaro et al. \cite{Catanzaro2005}.
Apparently this approach is used in many applications,
although sometimes no details
are given how these conflicts during the graph generation are solved,
like, e.g., in the original Ref.\ \cite{newman2001}.
Nevertheless, the ensemble generated in this way
exhibits a bias since now the sub-pairing probability depends
on the graph generated so far.
 This was observed for one sample degree sequence of a very small multigraph 
by King \cite{king2004}. The existence of this bias
 leads to the question, whether
this bias persists when going to larger, realistic graphs. In particular
it could be that for measurable quantities, within error bars, the
results of the true and the biased ensembles agree. We will
show in this paper that indeed the bias persists when
studying larger graphs, in particular it will be visible
for a global graph property, the graph \emph{diameter}.

The reminder of the paper is organized as follows:
Next, in the second section, we introduce a simple restricted ensemble, which
allows us to investigate its statistical properties
as a function of the system size. We will understand for its smallest
instance explicitly how the bias arises when repeating the creation of edges
in case of forbidden edges. In the third section, 
we study this bias as a function of the system size,
for a small range of sizes, where this is feasible. In the
fourth section, we investigate the behavior of the graph diameter
for ensembles of scale-free graphs.
Finally, in the conclusions, we summarize our results and discuss other
approaches like Markov chain Monte Carlo simulations \cite{taylor1982}
or a recently proposed rejection-free method \cite{delGenio2010} 
where the graphs carry additional weights.

\section{Example}

\label{sec:example}

\begin{figure*}
  \centering
    \includegraphics[width=0.9\textwidth]{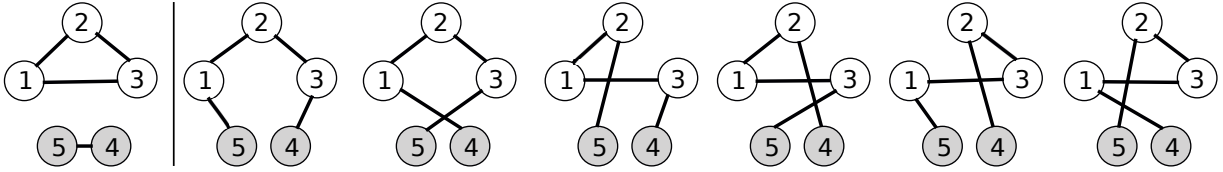}
  \caption{Seven possible realizations of a graph with five vertices 
for the degrees $d_1=2, d_2=2, d_3=2, d_4=1, d_5=1$}
  \label{fig:tops}
\end{figure*}

\begin{figure*}[htbp] \centering
    \includegraphics[width=0.9\textwidth]{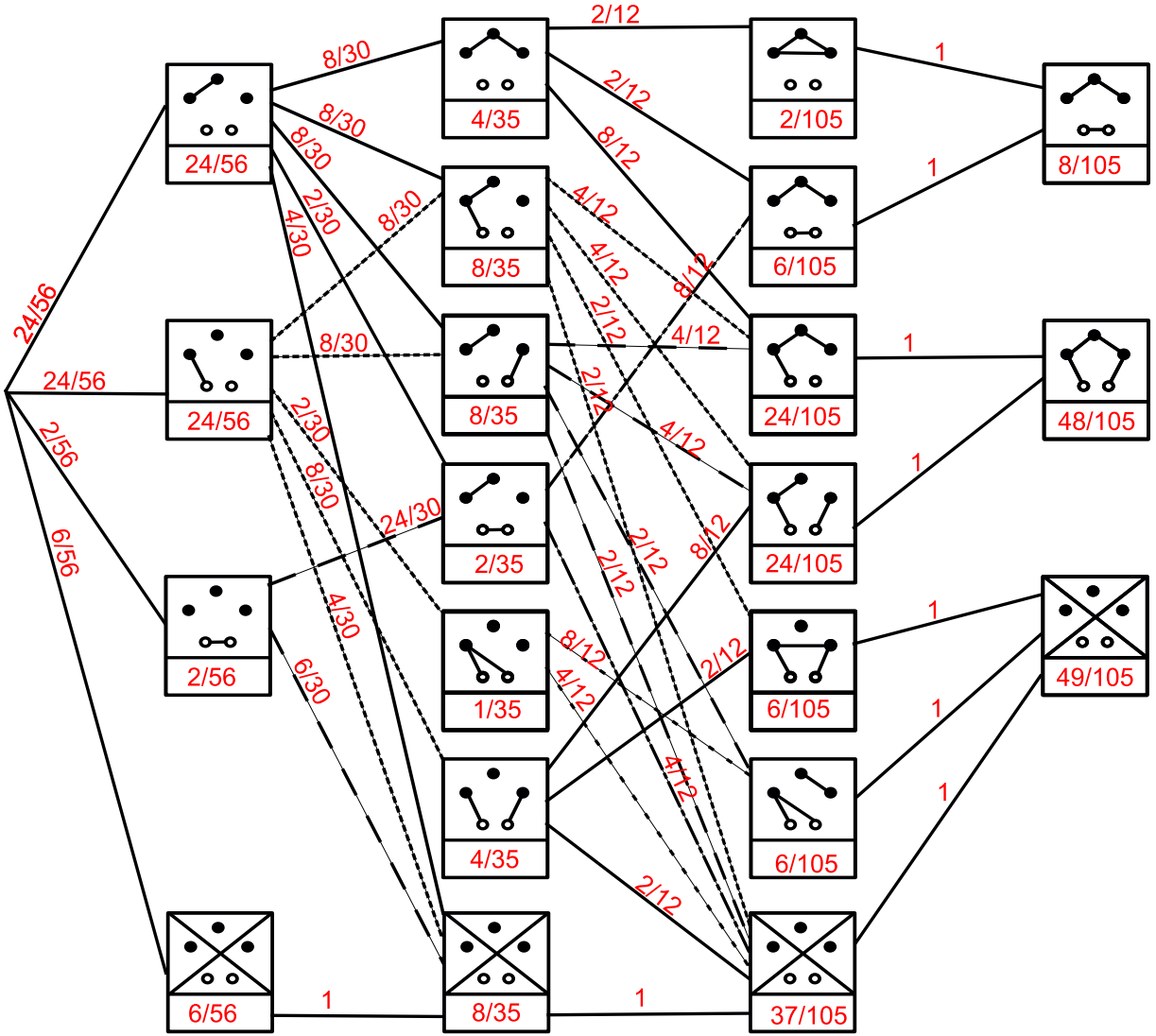}
  \caption{(color online) Generation of graph with five vertices having degrees
$d_1=2, d_2=2, d_3=2$ (black circles), $d_4=1, d_5=1$ (open
circles) with \emph{refusing} non-simple graphs. 
Starting from the left, in each step two stubs are picked
randomly and connected which can lead to several configurations,
shown by pictographs.
Configurations which are equivalent are summarized into one
pictograph.
For example, whether in the first step node 1 and 2, or node 1 and 3
are connected makes no difference. 
Possible transitions between configurations are indicated by lines.
The transition probabilities to reach a certain configuration
from the current one are  indicated by rational numbers shown next to
 the lines. The
probabilities to have reached a certain configuration are indicated by
the rational numbers beneath the pictographs. Reaching a non-valid
configuration is indicated by the ``crossed out'' graphs at the bottom
of the different columns.}
  \label{fig:stubs1}
\end{figure*}

In order to compare both variants of the configuration model, with edge
repetition or without (i.e., refusal),
we look at a simple example. Consider
a graph with 5 vertices. For this example and for the examples
in the subsequent section, we aim at graphs having roughly half of the
nodes exhibiting degree 1 and half of the nodes degree 2.
For our example here, the degree and therefore the number of stubs
for each vertex is fixed as follows: $d_1=2, d_2=2, d_3=2, d_4=1,
d_5=1$. Connecting all stubs can lead to two possible graph topologies
as shown in figure \ref{fig:tops}. Either vertex 4 and 5 are connected
and the resulting graph is split into two subgraphs ($A$), or the
graph is a single line with a variable order of the vertices
($B$). For an unbiased sampling, each of the seven realizations of the 
graph should occur with
the same probability. Since only one realization leads to the graph
topology $A$ and 6 to the graph topology $B$, the ratio
$\frac{p(A)}{p(B)}$ should be $\frac{1}{6}$. The process of building
up the graph by connecting stubs using the approach where the process
is restarted once an invalid edge is obtained (refusal)
is illustrated in figure \ref{fig:stubs1}.

One starts (left Fig.\ \ref{fig:stubs1}) with a completely unconnected graph.
In the first step
one out of eight stubs is picked, the stub is removed from the pool
and a second stub (out of the seven remaining ones) is picked at
random. The associated vertices are connected. There are four
different possible configurations after the first edge has been
made. Three valid edges are: Two vertices of degree 2 are
connected, a vertex with degree 2 is connected to a vertex with degree
1, or the two vertices with degree 1 are connected. A forbidden
edge is attempted, if the first and the second stub belong to
the same vertex, i.e., a self edge.
 This latter case is disregarded by restarting the
graph-generation process.  All probabilities shown in Fig.\ \ref{fig:stubs1}
can be easily calculated by hand.
 This leads finally to the probability $p$ for
the different topologies. The probability
to generate graph topology $A$ in one trial is $p(A)=\frac{8}{105}$,
$p(B)=\frac{48}{105}$ and the probability to end up disregarding the
graph is $p(C)=\frac{49}{105}$. Considering the valid topologies one
finds a ratio of the two graph topologies of $p(A)/p(B) = 1/6$ as
expected. Hence, even for this small graph for 
only about 50\% of the times  the graph-generation process
will lead to a valid graph
while in the other case the process has to be restarted from an empty graph.

\begin{figure*}[!ht] 
\centering
    \includegraphics[width=0.9\textwidth]{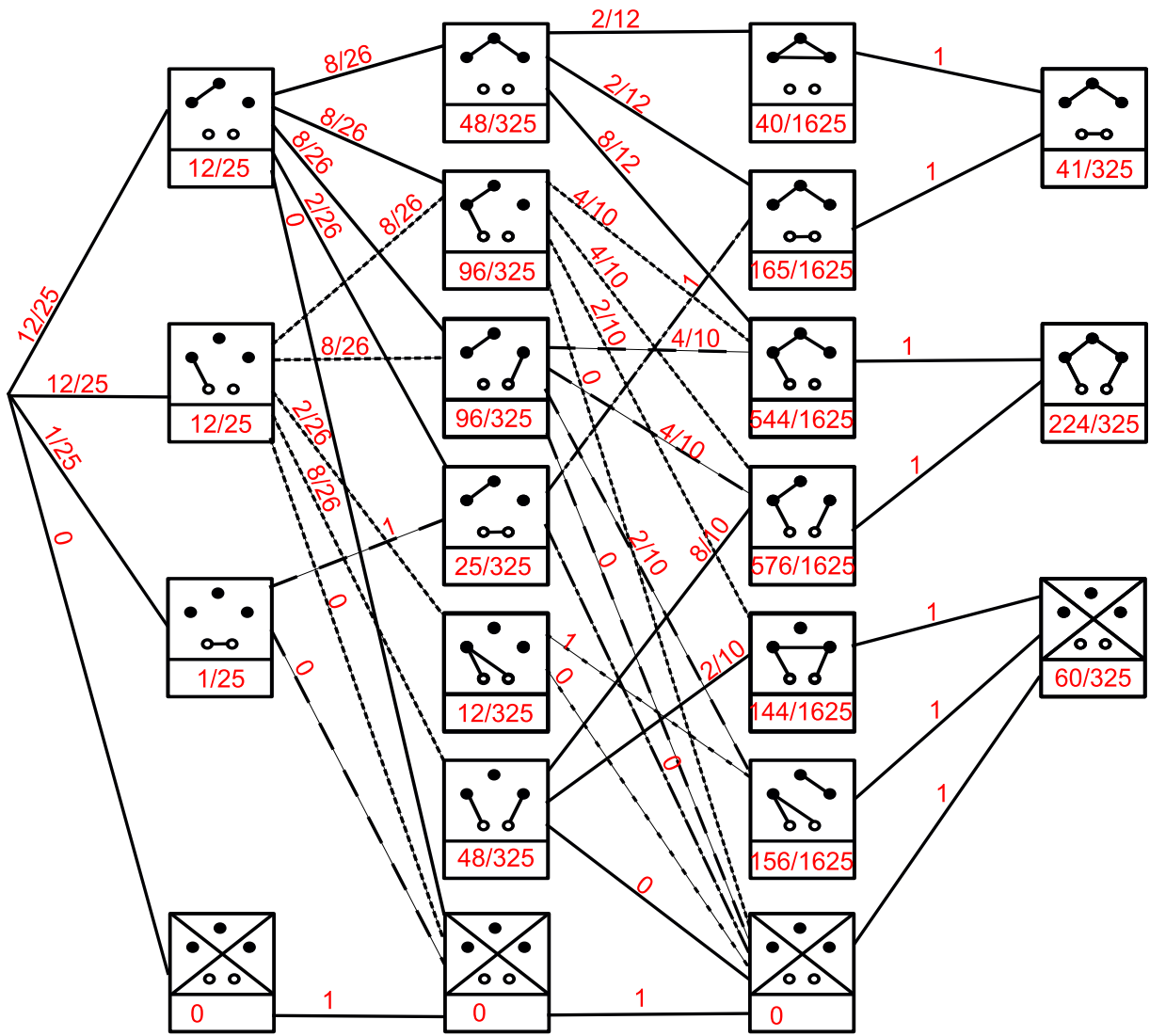}
  \caption{(color online) Generation of the same graph as shown
in Fig.\ \ref{fig:stubs1}, see there for details, but now
with \emph{repeated} selection of pairs of stubs in case invalid
edges are selected.
  \label{fig:stubs2}
}
\end{figure*}

Fewer restarts are needed, if the
second variant (\emph{repetition}) is applied,
i.e., if just the last pair of stubs leading to a forbidden
edge is disregarded. This generation process
is shown in figure \ref{fig:stubs2}. In each step the transition
probability $p_{\rm trans}(i\to j)$ from one configuration $i$ to another $j$ 
is calculated in the following way: It contains the probability
$p_{\rm direct}(i\to j)$ that the corresponding stubs are selected immediately.
Nevertheless, since after an invalid choice the step is repeated,
more terms contribute, e.g.,
the probability $p_{\rm error}(i)$ that first an invalid pair of stubs
is selected times the probability  $p_{\rm direct}(i\to j)$ that
 in the second try the corresponding stubs are selected. In the same
way, also the
probability contributes that two invalid tries are performed before a valid
pair of subs is selected, and so on:
\begin{eqnarray}
 p_{\rm trans}(i\to j) &=&\sum_{n=0}^{\infty}p_{\rm direct}(i\to j)
\left(p_{\rm error}(i)\right)^n\\
&= &\frac{p_{\rm direct}(i\to j)}{1-p_{\rm error}(i)}
\end{eqnarray}

By manually calculating these probabilities, on arrives at the
process displayed in Fig.\ \ref{fig:stubs2}. 
Note that as long as there exists a valid edge which can be
formed, this method does not have to restart the complete generation
process. For this reason, the probability that
one has to restart the full process is zero in early stages.
There are however configurations towards the termination of the process
which have only stubs left
that lead to forbidden edges. In these few cases the generation
process has to be restarted. The ratio of the two graph topologies is
$p(A)/p(B) = \frac{41}{224}$ which is different from the correct ratio
of $1/6=28/224$. The probability of forming a graph topology, where the graph
splits into two subgraphs is too high, compared to the formation of a
single line.

\section{Size-dependence}

\label{sec:sizes}

\begin{figure}[htbp]
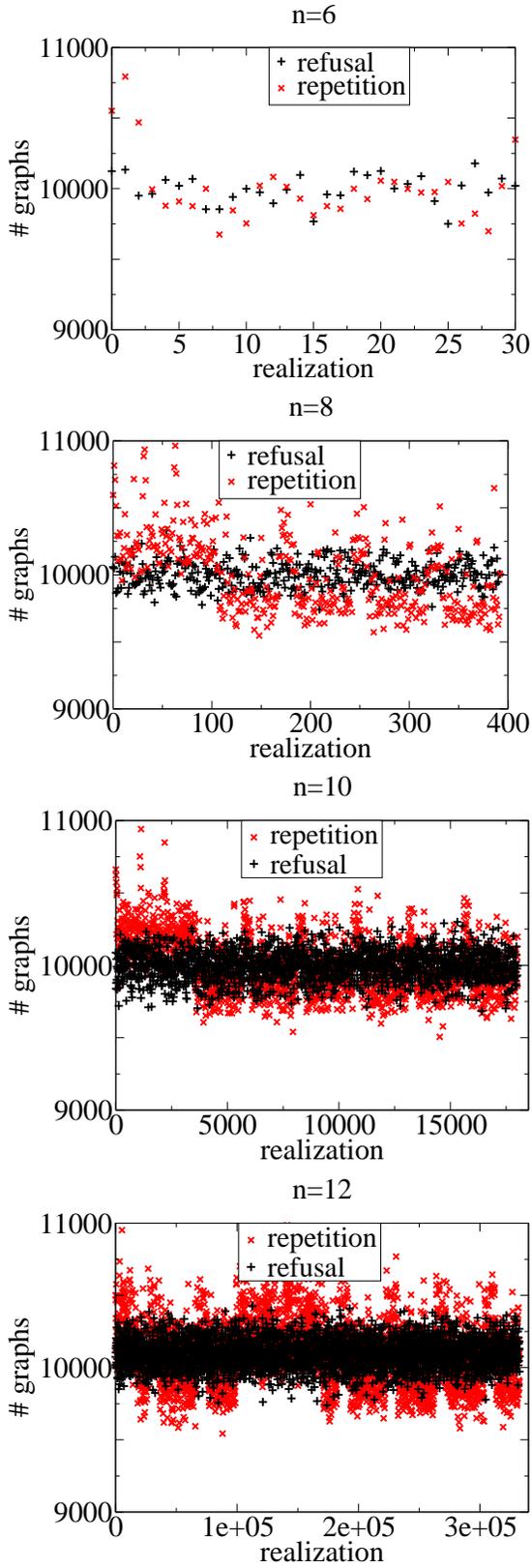
 \centering
    \includegraphics[width=0.4\textwidth]{n6_presentation.eps}

\vspace*{2mm}

    \includegraphics[width=0.4\textwidth]{n8_presentation.eps}

\vspace*{2mm}

    \includegraphics[width=0.4\textwidth]{n10_presentation.eps}

\vspace*{2mm}

    \includegraphics[width=0.4\textwidth]{n12_presentation.eps}
  \caption{(color online) 
Histograms for number of generated graph realizations. The two approaches
\emph{repetition} (red x symbols) and  \emph{refusal}
of non-simple graphs  (black $+$ symbols) are compared
 for different graph sizes $n=6,8,10,12$.
  \label{fig:stubshisto}
}
\end{figure}

The number of vertices and the degrees in the example in previous
section were chosen to be as simple as possible such that an effect
can be observed. Nevertheless, one might wonder whether for larger
graphs, the bias somehow decreases such that in the end it becomes
unimportant. In order to
investigate how this  effect behaves for larger $n$,
the two methods (refusal/repetition)
are compared numerically by
randomly generating a large number of graphs using a predetermined list of
stubs and counting the number of generated instances for each
graph realization. Again we aimed at graphs where roughly half of the
nodes have degree 1 and the other half has degree 2.
We considered graph sizes $n=6,8,10,12$. An even higher number
of nodes is not feasible, because the number of possible graphs
increases strongly.
 For $n=8$ and $n=12$ exactly half of the vertices have degree 1
 and the other half has 
degree 2. For  $n=6$ ($n=10$) two (four) vertices have
degree 1 and four (six) vertices degree 2. This results in an even
number of stubs in all cases. The number of generated graphs was
chosen such that on average 10000 graphs were created per possible
graph realization. The exact number of possible realizations (bins of the
histogram) for each graph size $n$
 can be found in table \ref{tab:chisq}.
The resulting histograms are shown in figure \ref{fig:stubshisto}.

For all sizes shown in figure \ref{fig:stubshisto} each realization
occurs, within statistical variation, with the same probability, if
the entire graph is refused as soon as the first forbidden edge
occurs. If just the forbidden edges are disregarded (repetition),
some realizations appeared significantly
 more often than others. These deviations are
especially prominent for large graph sizes. Hence, the bias
of the repetition method does not disappear!

The make the statement also more quantitatively, we calculated
 \emph{p-values} from a
chi-squared test \cite{practical_guide2009}
for the sampled histograms assuming an equal
distribution of realizations. The resulting p-values are
 shown in table \ref{tab:chisq}. In
addition to the p-values from 10000 sampled graphs per bin as used in the
figures, also the resulting
 p-values for a smaller number of 1000 graphs per bin are shown.

\begin {table} [ht]
	\begin{tabular}{rrcrccrcc}
	\hline 
      {n}	& {\# bins}	& \vline	& \multicolumn{2}{c}{{1000/bin}}			& \vline	& 
\multicolumn{2}{c}{{10000/bin}} 		\\
			&  			& \vline	& {refusal}	& {repetition}& \vline	& {refusal}& {repetition}\\
	\hline
	 6 	& 31	& \vline	& $0.75$	& $2.8\cdot10^{-3}$	& \vline	& $0.33$	& $4.6\cdot10^{-23}$	\\
	 8 	& 393	& \vline	& $0.52$	& $8.5\cdot10^{-19}$	& \vline	& $0.76$	& $0.0$			\\
	 10	& 18012 & \vline	& $0.15$	& $4.1\cdot10^{-142}$	& \vline	& $0.36$	& $0.0$ 		\\
	 12	& 332790& \vline	& $0.91$ 	& $0.0$			& \vline	& $1.0$ 	& $0.0$     		\\
	\hline	
	\end{tabular}
\caption{The p-values obtained from a chi-squared test for the 
sampled histograms assuming an uniform distribution of realizations 
($\leftrightarrow$ bins) for 
an average of 1000 and 10000 sampled graphs per bin.}
\label{tab:chisq}
\end {table}

The p-value for the refusal method 
varies  between 0.15 and 1.00 for the
different cases. This statically supports the result that all
graph realizations are equiprobable, hence the ensemble is not biased.
When using the method with repetition of edge creation the
value is by several orders of magnitude smaller and decreases quickly
for increasing graph size $n$. For a large number of samples per bin and/or
large graphs, the p-value is even just 0 within the standard numerical 
accuracy. This clearly shows that the different graphs are not
equiprobable.

Another way to analyze the statistics
of the graph generation process is to find a measurable quantity that
might differ on average when using different generation methods. 
Here, we considered the graph diameter , which is among all
pairwise shortest path distances of a graph the longest one (omitting
infinite distance if two nodes are not connected). 
For the sample graphs of Sec.\ \ref{sec:example}, one obtains
an average diameter of $212/156\approx 3.79$ for the (unbiased) refusal
approach, while for the repetition approach a much smaller value
of $978/325 \approx 3.01$ is obtained. For larger systems
sizes the difference becomes smaller, but still measurable:
The average diameter $d_{\max}$ and the standard error
for $10^7$ randomly generated graphs of size $n=6,8,10,12$ with degrees
as described above can be found in
table \ref{tab:diameter}.
We state differences  of the diameters normalized by the
maximum standard error $\tilde \sigma$ given by
\begin{equation}
\Delta=(d_{\max}({\rm refusal}) -d_{\max}({\rm repetition}))/\tilde \sigma\,.
\label{eq:difference}
\end{equation}
The differences in diameter are now quite small but still 
statistically significant.
Hence, studying measurable quantities might be a promising approach
to investigate ensembles of even larger graphs, which we present in the 
next section.

\begin {table} [h]
	\begin{tabular}{ccccccc}
	\hline 
        $n$ &\vline& $d_{\max}$ refusal &\vline&  
$d_{\max}$ repetition & \vline& $\Delta$ \\
	\hline
	 6  & \vline& $2.4839 \pm 0.0001$ & \vline& $2.4771 \pm 0.0001$ & \vline&68\\
	 8  & \vline& $2.4580 \pm 0.0001$ & \vline& $2.4570 \pm 0.0001$ & \vline&10\\
	 10 & \vline& $2.8632 \pm 0.0001$ & \vline& $2.8628 \pm 0.0001$ & \vline&4 \\
	 12 & \vline& $2.7831 \pm 0.0001$ & \vline& $2.7843 \pm 0.0001$ & \vline&-12\\
	\hline	
	\end{tabular}
\caption{Mean diameter $d_{\max}$
and standard error for $10^7$ sampled graphs, for the \emph{refusal} and
\emph{repetition} approaches, respectively, for different graph sizes $n$.
 The last column displays the normalized differences $\Delta$
 of the diameters.}
\label{tab:diameter}
\end {table}

\section{Diameter of scale-free graphs} 
\label{sec:scale-free}

\begin{figure}[htbp]
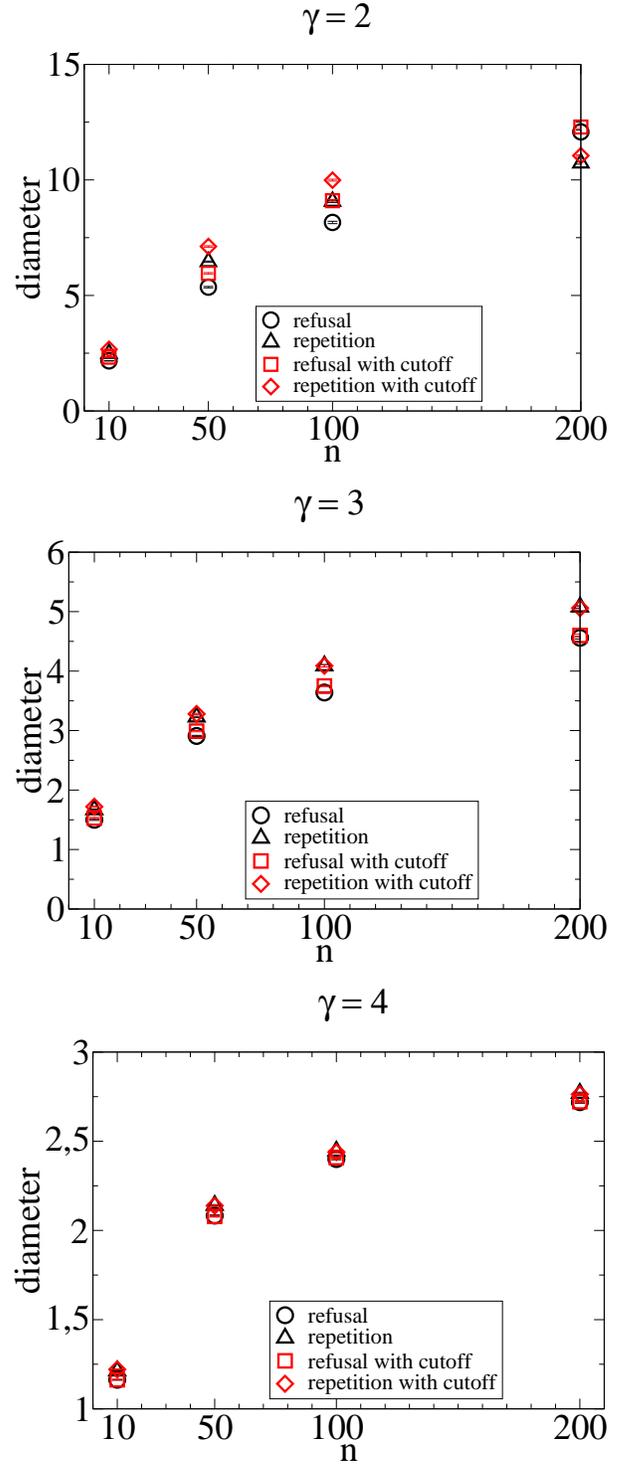

  \centering
    \includegraphics[width=0.44\textwidth]{diameters_g2.eps}

\vspace*{3mm}

    \includegraphics[width=0.44\textwidth]{diameters_g3.eps}

\vspace*{3mm}

    \includegraphics[width=0.44\textwidth]{diameters_g4.eps}
  \caption{(color online) Average diameter $d_{\max}$ 
of scale-free graphs with exponents of $\gamma=2$, $3$ 
and $4$ for $10^4$ generated graphs as a function of the number 
of vertices $n$ using four different
approaches. The error bars are smaller than the symbol size.}
  \label{fig:diameter}
\end{figure}

To investigate, whether the bias present in the repetition approach
is measurable for even larger graphs, we studied
scale-free graphs where the node degrees are sampled from
the distribution shown in Eq.\ \ref{eq:scale-free}.
For $\gamma\leq2$ the first moment of this distribution
diverges in the thermodynamic limit. For $\gamma\leq3$ the second 
moment diverges. This can lead to vertices with very high degrees. Figure
\ref{fig:diameter} shows the mean diameter of scale-free graphs with
exponents of $\gamma=2$, $3$, $4$ for 10000 generated graphs as a
function of the number of vertices $n$.
Here we also generated graphs where the maximum degree is cut off at
$\sqrt{n}$.
  This cutoff was suggested by Catanzaro et al. \cite{Catanzaro2005}
in order to remove degree correlations, which occur otherwise for
$\gamma<3$.  Also preventing very high degrees allows to generate 
larger graphs
at low values of $\gamma$ using the refusal approach.
Nevertheless,  we only studied $\gamma \ge 2$, because otherwise the
refusal method does not allow to study very large systems for all
cases.
 
For $\gamma=2$ the
diameters deviate depending on the method which is used for graph
creation. The deviation increases for increasing system size. Including
the cutoff increases the diameters for both generation approaches, 
but it does not reduce
the difference between both generation methods.
For $\gamma=3$, the differences between the different case become
smaller. For $\gamma=4$ there is even 
very little difference in diameter for graphs
generated with both methods. Introducing a cutoff for the highest
degree does not change the diameter at all in this case, since high
degrees are anyway very rare.

To study the differences between the unbiased refusal and the biased repetition
approaches more quantitatively, we also calculated the
normalized differences $\Delta$ (Eq. \ref{eq:difference}) 
between the average diameters, see Tan.\ \ref{tab:errors}.
Clearly, even for large graphs, the error introduced
by the bias of the repetition approach is quite significant.

\begin {table} [ht]
	\begin{tabular}{ccccccccc}
	\hline 
   $\Delta$         &\vline& {$n=10$} &\vline&  {$n=50$} & \vline& {$n=100$} & \vline & {$n=200$} \\
	\hline
	 $\gamma=2$  & \vline& -35 & \vline& -37 & \vline & -18 & \vline & 17 \\
	 $\gamma=2$ cutoff    & \vline& -34 & \vline& -39 & \vline & -18 & \vline & 21 \\
	 $\gamma=3$  & \vline& -28 & \vline& -32 & \vline & -23 & \vline & -26 \\
	 $\gamma=3$ cutoff    & \vline& -32 & \vline& -29 & \vline & -17 & \vline & -15 \\
	 $\gamma=4$  & \vline& -25 & \vline& -15 & \vline & -11 & \vline & -8 \\
	 $\gamma=4$ cutoff    & \vline& -29 & \vline& -15 & \vline & -8 & \vline & -7 \\
	\hline	
	\end{tabular}
\caption{Normalized difference $\Delta$ between the average diameters
for scale free graphs ($\gamma=2,3,4$), also when additionally introducing
a degree cutoff at $\sqrt{n}$.
\label{tab:errors}}
\end {table}

\section{Conclusions}

We have studied the statistics of generating random (labeled, simple) 
undirected graphs
with prescribed degree sequences using the configuration-model
approach. The basic idea is to assign each node a number of stubs
equal to its degree and then pair randomly selected stubs.
This leads to an unbiased sample, i.e., the weight is uniform,
if only valid simple graphs are kept and all others refused.
In many cases, e.g., for scale-free graphs with a very broad tail,
this is very inefficient. Hence, in the literature very often an approach
is used where invalid edges
are immediately rejected and instead the current stub-selection
step is repeated. In this way much larger graphs can be generated,
even for a very broad degree distribution. Nevertheless, this ensemble
is biased, as can be seen from a very simple example of a graph
with five nodes.

Our statistical analysis of related degree sequences for graphs
sizes $n=6$, 8, 10, 12 shows that this bias does not disappear.
Instead, by using a chi-squared test, we could show that the
p-value decreases quickly such that it becomes zero within
the numerical accuracy. Even worse, for measurable quantities like
the diameter, the average estimates differ by many error bars,
also for much larger sizes such as $n=200$. Hence, for a careful
analysis of graphs, either with prescribed degree sequences or for an
ensemble obtained by randomly drawing degree sequences, the
configuration model with repetition does not work properly.

Recently, Del Genio et al.\ \cite{delGenio2010} proposed a rejection-free
approach which is based on restricting the sampling in each step
to the set of edges such that for the remaining degree sequence there
is still a simple undirected graph. The algorithm introduces a bias
which is controlled by calculating weights which allow for correcting
for this bias. Unfortunately, the distribution of weights exhibit a log-normal
distribution. Hence, in each set of sampled graphs, there will be few 
 samples which nevertheless carry a weight which is many orders
of magnitude larger than the typical weight. In a sample
study in Ref. \cite{delGenio2010} for scale-free graphs with $\gamma=3$
and $n=100$, the largest sampled graph (among $10^6$ graphs) 
exhibited a weight which exceeded
the typical weight by a factor of $10^{26}$. Hence, to calculate any measurable
quantity, an extremely small number of graphs (usually one if the graphs 
are large) will contribute within a set of graphs
generated by this approach.
Note that the algorithm of Blitzstein and Diaconis \cite{blitzstein2006},
which was proposed in 2006, suffers from the same problem. Therefore, it
would be very useful to alter these approaches such that preferentially
graphs with large weight are generated. Whether this is possible
remains an open question in the moment.

Hence, to our knowledge, there is to data no practical approach which allows
efficiently  to    sample graphs rejection-free of any given degree sequence
without a bias. Hence, it appears
that one still should use methods which are based on Markov-chain
Monte Carlo methods, i.e. which start with any feasible graph 
and perform swaps of randomly selected
edges \cite{taylor1982}. Unfortunately, this approach creates
correlations between subsequent graphs, hence one has to take additional
effort to estimate mixing (decorrelation) times.

\section{Acknowledgements}

We thank Charo Del Genio for useful discussions.
The simulations were performed at the HERO high-performance
computing facility (supported by the DFG, INST 184/108-1 FUGG)
 at the University of Oldenburg. 

\bibliographystyle{apsrev}
\bibliography{biased_config_model.bib}

\end{document}